\documentclass[superscriptaddress,twocolumn,notitlepage,aps,prl,preprintnumbers,nobibnotes]{revtex4-2}

\usepackage{amsmath}
\usepackage{physics}
\usepackage{amsfonts}
\usepackage{graphicx}
\usepackage[dvipsnames]{xcolor}
\usepackage{qcircuit}
\usepackage{dsfont}
\usepackage{xfrac}
\usepackage{xr}
\usepackage{enumitem}
\usepackage{mathtools}
\usepackage{tikz}
\usepackage{dsfont}

\usetikzlibrary{shapes}

\newcommand{\thetitle}{Error mitigation via stabilizer measurement emulation}
\newcommand{\comment}[1]{}

\newcommand{\beq}{\begin{equation}}
\newcommand{\eeq}{\end{equation}}

\newcommand{\SQM}{\textsf{QME}}
\newcommand{\QME}{\textsf{QME}}
\newcommand{\CZ}{\textsf{CZ}}

\newcommand{\Id}{\mathds{1}}

\newcommand{\RLEaffil}{Research Laboratory of Electronics, Massachusetts Institute of Technology, Cambridge, MA 02139, USA}
\newcommand{\EECSaffil}{Department of Electrical Engineering \& Computer Science, Massachusetts Institute of Technology, Cambridge, MA 02139, USA}
\newcommand{\LLaffil}{MIT Lincoln Laboratory, Lexington, MA 02421, USA}
\newcommand{\CHALMERSaffil}{Microtechnology and Nanoscience, Chalmers University of Technology, G\"oteborg, SE-412 96, Sweden}
\newcommand{\MITPHYSaffil}{Department of Physics, Massachusetts Institute of Technology, Cambridge, MA 02139, USA}
\newcommand{\MITMECHaffil}{Department of Mechanical Engineering, Massachusetts Institute of Technology, Cambridge, MA 02139, USA}
\newcommand{\DUKEaffil}{Departments of Physics \& Electrical and Computer Engineering, Duke University, Durham, NC 27708, USA}

\begin{document}
\date{\today}
\begin{abstract}
Dynamical decoupling (DD) is a widely-used quantum control technique that takes advantage of temporal symmetries in order to partially suppress quantum errors without the need resource-intensive error detection and correction protocols. This and other open-loop error mitigation techniques are critical for quantum information processing in the era of Noisy Intermediate-Scale Quantum technology. However, despite its utility, dynamical decoupling does not address errors which occur at unstructured times during a circuit, including certain commonly-encountered noise mechanisms such as cross-talk and imperfectly calibrated control pulses. Here, we introduce and demonstrate an alternative technique --- `quantum measurement emulation' (QME) --- that effectively emulates the measurement of stabilizer operators via stochastic gate application, leading to a first-order insensitivity to coherent errors. The QME protocol enables error suppression based on the stabilizer code formalism without the need for costly measurements and feedback, and it is particularly well-suited to discrete coherent errors that are challenging for DD to address.
\end{abstract}

\title{\thetitle}

\author{A.~Greene}\affiliation{\RLEaffil}
\author{M.~Kjaergaard}\affiliation{\RLEaffil}
\author{M.~E.~Schwartz}\affiliation{\LLaffil}
\author{G.~O.~Samach}\affiliation{\RLEaffil}\affiliation{\LLaffil}
\author{A.~Bengtsson}\affiliation{\RLEaffil}\affiliation{\CHALMERSaffil}
\author{C.~McNally}\affiliation{\RLEaffil}
\author{M.~O'Keeffe}\affiliation{\LLaffil}
\author{D.~K.~Kim}\affiliation{\LLaffil}
\author{M.~Marvian}\affiliation{\RLEaffil}\affiliation{\MITMECHaffil}
\author{A.~Melville}\affiliation{\LLaffil}
\author{B.~M.~Niedzielski}\affiliation{\LLaffil}
\author{A.~Veps\"{a}l\"{a}inen}\affiliation{\RLEaffil}
\author{R.~Winik}\affiliation{\RLEaffil}
\author{J.~Yoder}\affiliation{\LLaffil}
\author{D.~Rosenberg}\affiliation{\LLaffil}
\author{S.~Lloyd}\affiliation{\RLEaffil}\affiliation{\MITMECHaffil}
\author{T.~P.~Orlando}\affiliation{\RLEaffil}
\author{I.~Marvian}\affiliation{\DUKEaffil}
\author{S.~Gustavsson}\affiliation{\RLEaffil}
\author{W.~D.~Oliver}\affiliation{\RLEaffil}\affiliation{\LLaffil}\affiliation{\MITPHYSaffil}\affiliation{\EECSaffil}

\maketitle

\section*{Introduction}

Over the past few years, impressive strides have been made in the field of quantum computing. Quantum advantage has been reported \cite{google_supremacy_2020} and there is now an ecosystem of cloud-based quantum processors and companies interested in using them \cite{henderson_space_2020, chow_scalable_2015, rigetti_instructionset_2017, ibm_cloud_error_mitigation_2020}. However, high error rates continue to limit circuit depth such that solving real-world problems with today's quantum computers remains a challenge \cite{martin2019pricing, zapata_chem_review_2019}. Errors in quantum devices can be grouped into two categories: incoherent (stochastsic) and coherent (deterministic). Although coherent errors do not reduce the purity of a qubit state, they are surprisingly pernicious. For small errors, infidelity grows linearly with incoherent errors and quadratically with coherent errors \cite{unitarity_2019}. As quantum processors grow in complexity and computational power, control signal cross-talk and coherent errors from spectator qubits become increasingly problematic for gate performance and increasingly challenging to characterize \cite{wallraff_spectatorqb_2020}. Unlike incoherent errors, they compound each other in unpredictable ways \cite{wallman_coherenterr_2015} that lead to substantially worse performance than predicted by randomized benchmarking techniques.

Feedback-based quantum error correction (QEC) is designed to suppress the high error rates in quantum hardware, and attaining QEC is a long-term goal of the quantum computation community \cite{qec_book, xzzx_2020, heavyhex_2020, ionqecc_2020, catcode_2020}. QEC works by making error detection measurements that are carefully designed to preserve quantum information, and then performing restorative gates on the system conditioned on the error pattern observed in order to return the processor to an error-free state. However, performing feedback is an enormous technical challenge. Compared to other operations on today's superconducting qubit quantum processors, readout has the highest error rate by an order of magnitude \cite{google_supremacy_2020, jurcevic_2020, stehlik_2021}. Readout is also the longest operation on these processors by up to two orders of magnitude \cite{Sank_2016, stehlik_2021}, so teams working to get a feedback operation below the threshold for fault-tolerant computation must wrestle with a substantial incoherent error as well as a relatively low assignment fidelity \cite{riste_feedback_2013}. It thus behooves us to get as much mileage as we can out of error reduction techniques which do not utilize feedback (\textit{i. e.} feed-forward error mitigation), especially in the era of Noisy Intermediate-Scale Quantum (NISQ) technology \cite{Preskill_2018}. 

Dynamical decoupling (DD) is the primary feed-forward control technique in use today \cite{PhysRevA.58.2733, Khodjasteh_2005, Zanardi_1999, Viola_1999, Uhrig_2007}. It takes advantage of temporal symmetries in the structure of the noise in order to reverse errors, and is highly effective against low frequency incoherent noise. The canonical example of such a technique is the Hahn echo, pioneered in the context of magnetic resonance imaging \cite{hahn_1958}. However, most sources of coherent error (such as imperfectly calibrated gates and cross-talk) are discrete events with no inherent temporal symmetry to exploit.  

Randomized compiling \cite{wallman_randcomp_2016, rand_compile_expt_2020} is a promising new feed-forward control technique that specifically addresses coherent errors from complex gates by inserting stochastically chosen high-fidelity gates to make those errors incoherent. A twirling gate $T_k$, drawn uniformly at random from the group generated by Pauli gates and the phase gate, is inserted before each error-prone gate $G_k$. To preserve the original computation, $G_k$ is followed by a correction gate  $T^c_k$ calculated such that $T^c_k G_k T_k = G_k$. This renders coherent errors incoherent, which reduces error rates and lowers the gate fidelity threshold required for fault-tolerant quantum computation.         

In this work, we introduce a related feed-forward control technique for addressing coherent errors, which uses stochastic gates but requires no classical pre-computation. Previous work on stabilizer codes has shown that errors can be reduced simply by measuring the stabilizers, even without feedback \cite{lidar_dd_2018, mcclean_subspaceexpansion_2019, muller_quantumzeno_2019, kempf_zeno_2015, Facchi_2004} \comment{[Lidar paper]}. We take this a step further and investigate the effects of ‘simulating’ the quantum channel of measurement via the stochastic application of single-qubit gates, which is faster than real measurement by more than an order of magnitude. We find that this technique leads to an improvement in circuit performance. This ``quantum measurement emulation" ($\SQM$) technique extends naturally from the stabilizer code formalism and is particularly well-suited for discrete coherent errors that are challenging for dynamical decoupling to address.

\section*{Operating Principle of ‘Emulating’ A Quantum Measurement Channel}
 
  Consider the measurement of a state $\rho$ along the $z$-axis in an experiment without feedback, described by the Kraus operators $P^z_0 = |0\rangle\langle0|$ and $P^z_1 = |1\rangle\langle1|$:
\begin{equation}
\begin{split}
\mathcal M_z(\rho): \rho \mapsto \sum_{i=0,1} P^z_i\rho P^{z\dagger}_i \\
\mathcal M_z
\left(
\begin{bmatrix} 
\rho_{11} & \rho_{12} \\
\rho_{21} & \rho_{22} \\
\end{bmatrix}
\right)  = 
\begin{bmatrix} 
\rho_{11} & 0 \\
0 & \rho_{22} \\
\end{bmatrix}
\end{split}
\label{eq:measKraus}
\end{equation}

The same quantum channel (i.e. mapping) is implemented by

\begin{equation}
\begin{split}
\mathcal D_z(\rho): \rho \mapsto \frac{1}{2}\Id\rho\Id + \frac{1}{2}Z\rho Z \\
\mathcal D_z \left(
\begin{bmatrix} 
\rho_{11} & \rho_{12} \\
\rho_{21} & \rho_{22} \\
\end{bmatrix}
\right) = 
\begin{bmatrix} 
\rho_{11} & 0 \\
0 & \rho_{22} \\
\end{bmatrix}
\end{split}
\label{eq:dephasing}
\end{equation}

\noindent which is the dephasing operator \cite{mikenike}. Equation \eqref{eq:dephasing} provides an operational interpretation of the act of measurement: the state is dephased in the basis of the measurement. Equation \eqref{eq:dephasing} can also be interpreted as a probabilistic application of either the identity operation ($\Id$) or a Pauli $Z$ operation.

\begin{figure}[!ht]
	\center
	\includegraphics[width=1\columnwidth]{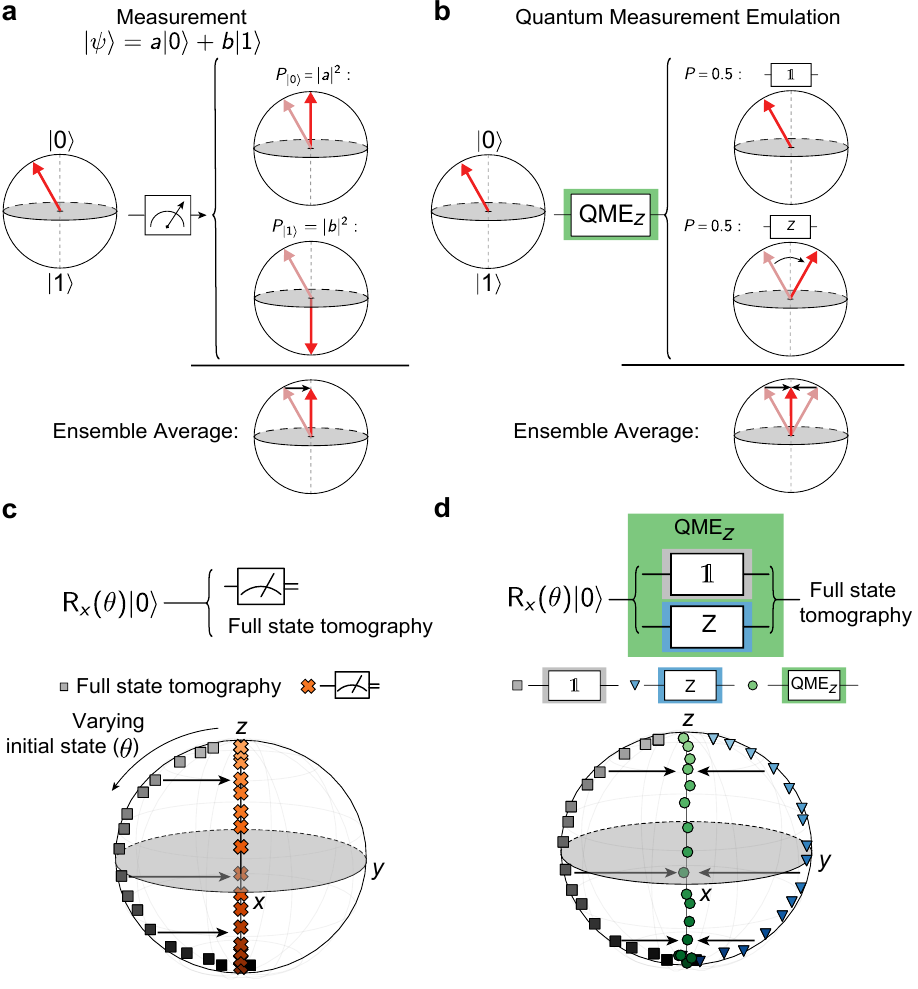}
	\caption{\footnotesize{\textbf{Emulated dephasing channel}
			\textbf{a} Diagram depicting $z$-axis measurement as a dephasing channel. The measurement projects the quantum state onto the $z$-axis and reduces the length of the Bloch vector. 
			\textbf{b} Stochastic gate application can create the same quantum channel as $z$-axis measurement.
			\textbf{c} Various qubit initial states reconstructed with state tomography (grey squares), plotted with projection after $z$-axis measurement (orange x).
			\textbf{d} State tomography data of qubit $\rho$, plotted after applying $\Id$ (grey squares), Pauli $Z$ gate (blue triangles) and $\SQM_Z$ (green circles) for varying initial states. We see that $\SQM_Z$ implements the same channel as $z$-axis measurement. 
	}}
	\label{fig:SQM_cartoon}
	
\end{figure}
Taken together this observation implies that when averaging over an ensemble of identically prepared qubits, applying a $Z$ gate with probability 0.5 (and otherwise doing nothing) has the same effect on the qubit’s density matrix $\rho$ as performing a measurement in the $z$ basis. We call this stochastic operation \textit{quantum measurement emulation}:

\begin{equation}
\Qcircuit @C=0.5em @R=1em{
	&\gate{\SQM_Z} & \qw\\
}
=
\begin{cases}
\Qcircuit @C=0.5em @R=1em{ &\gate{\Id} & \qw \\} & \text{with } p = \frac{1}{2}\\

\Qcircuit @C=0.5em @R=1em{&\gate{Z} & \qw \\} & \text{with } p = \frac{1}{2}
\end{cases}
\label{eq:sqm_def}
\end{equation}

For an intuitive picture, consider a vector on the Bloch sphere. Fig 1a visualizes the effect of $z$-axis measurement when performed on an ensemble of identically prepared experiments. In Fig 1b, we see how applying the $\SQM$ operation has the same outcome ---  when the $\Id$ and $Z$ branches are averaged together, their $x$- and $y$- components cancel and leave only the projection on the $z$-axis.

We experimentally demonstrate the equality of these two quantum channels using superconducting transmon qubits. (For more information on the device, see \cite{kjaergaard_dme_2020}). We prepare the qubit in an arbitrarily chosen initial state and plot a tomographic reconstruction of the state after either doing nothing (grey square), applying a Z gate (blue triangle), or after stochastically applying a $Z$ gate with p = 0.5 (green circle). We repeat these measurements for a variety of input states. As expected from Equation \eqref{eq:dephasing}, the probabilistic application of $Z$ and $\Id$ gates effectively projects the qubit state onto the $z$-axis, with some small deviations due to measurement errors.  To highlight the effective equivalence between this and actual $z$-basis measurement, we prepare the same initial states and plot $\expval{Z}$ (Fig 1c).  

The equivalence we have described between $\mathcal M_z(\rho)$ and $\mathcal D_z(\rho)$ can be generalized to measurement along any axis, including those in a multi-qubit basis. The observable for an $n$-qubit measurement along a multi-qubit axis $\nu$  corresponds to a unitary $\mathcal S_\nu \in SU(2^n)$. Measurement along $\nu$ is performed by replacing the projectors in Equation \eqref{eq:measKraus} with the projectors for the $\pm 1$ eigenstates of  $\mathcal S_\nu$. Consequently, the generalized dephasing channel is given by:

\begin{equation}
\mathcal D_\nu(\rho): \rho \mapsto \frac{1}{2}\Id^{\otimes n}\rho\Id^{\otimes n} + \frac{1}{2}\mathcal S_\nu \rho \mathcal S_\nu^\dag
\label{eq:generaldephasing}
\end{equation}
This implies that a quantum measurement emulation of any Pauli operator can be performed with the stochastic application of single-qubit gates. As an example, we show below how the quantum channel for $\expval{ZZ}$ measurement can be emulated using single-qubit $Z$ gates. 

\begin{equation}
\hspace*{-1cm}
\raisebox{0.45cm}{
\Qcircuit @C=0.5em @R=1em{
	&\multigate{1}{\SQM_{ZZ}} & \qw \\
	&\ghost{\SQM_{ZZ}} & \qw \\
}}
=
\begin{cases}
\raisebox{0.45cm}{
\Qcircuit @C=0.5em @R=1em{
	&\gate{\Id} & \qw \\
	&\gate{\Id} & \qw \\
}} & \text{   with } p = \frac{1}{2}\\
\\
\Qcircuit @C=0.5em @R=1em{
	&\gate{Z} & \qw \\
	&\gate{Z} & \qw \\
}& \raisebox{-0.45cm}{\text{with } \it{p} = $\frac{1}{2}$}
\end{cases}
\label{eq:sqm_def}
\end{equation}
An important example of this is when $\mathcal S_\nu$ is the stabilizer for a logical qubit. Stabilizer codes are designed such that stabilizer measurements do not corrupt the quantum information stored in the logical qubit, and the measurement results can be used as a parity check for error detection or error correction. In this case, performing $\SQM_\nu$ emulates a stabilizer measurement which can be used to mitigate coherent errors which rotate the state outside of the codespace.

\section*{Using $\SQM$ to Mitigate Coherent Rotation Errors}

\begin{figure}[!ht]
	\center
	\includegraphics[width=\columnwidth]{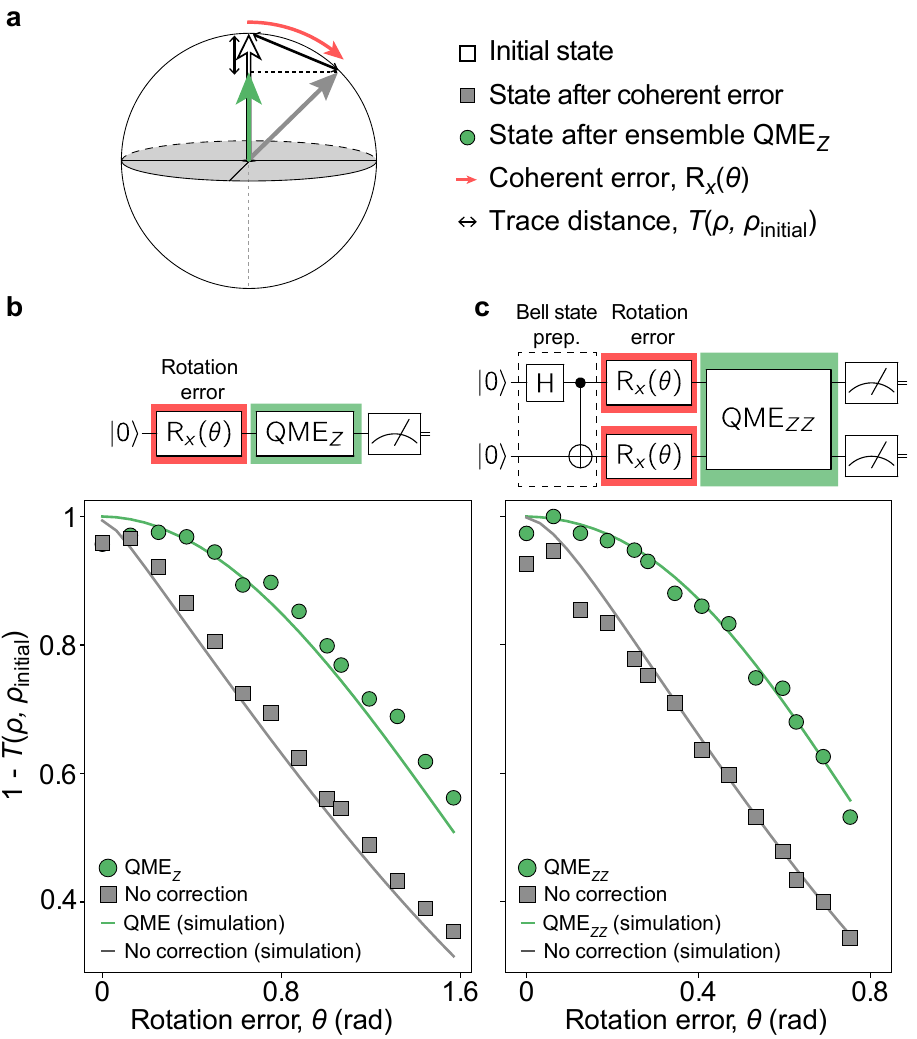}
	\caption{\textbf{$\SQM$ for Error Mitigation}
		\footnotesize{
		 	\textbf{a} Diagram depicting $\SQM$ as an error mitigation technique. After  the  initial  state  is prepared (white circle), a coherent error rotates  the  state  away (grey  square). Applying $\SQM_Z$ projects that error into an incoherent error, pushing the qubit state back to the $z$-axis (green circle). Notice  that  after  performing  $\SQM$, the  trace  distance to the initial state is greatly reduced.
		 	\textbf{b} Qubit is prepared in the $\ket{0}$ state and subjected to coherent errors of increasing strength. We plot the trace distance with (grey squares) and without (green circles) $\SQM$, overlaid with theoretical values from simulations performed using the measured initial state and no free parameters.
		 	\textbf{c} Qubits are prepared in $\overline{\ket{0}} = \frac{1}{\sqrt{2}}(\ket{00} + \ket{11})$ in a simple code stabilized by $ZZ$, subjected to a small $R_x(\theta)\otimes R_x(\theta)$ coherent error and plot the trace distance to the initially prepared state with (green circles) and without (grey squares) $\SQM_{ZZ}$ for increasing error strength $\theta$. In both of these examples, $\SQM$ provides a first-order insensitivity to $\theta$. (In these plots, $1-T(\rho, \rho_{initial})$ has been normalized to 1 in order to account for state preparation and measurement (SPAM) errors.) 
		}
	}
	\label{fig:SQM_cartton}
		
\end{figure}

After a coherent error has occurred, applying the appropriate $\SQM$ will project the qubit onto the measurement axis, effectively converting the coherent error into an incoherent error. Fig. 2a shows a qubit initially in the $\ket{0}$ state (white) which is subjected to a coherent rotation error $R_x(\theta)$ (grey). When $\QME$ is performed along the $z$-axis (denoted $\SQM_Z$), the qubit state is projected onto the $z$-axis. After $\SQM$, the coherent error which rotated the qubit state on the surface of the Bloch sphere is now manifest as an incoherent error which shortens the length of the vector along the measurement axis. After applying $\SQM_Z$ the trace distance between the initial and measured states is only second order in the size of the original perturbation. Here, trace distance $T(\rho, \sigma) = \frac{1}{2} \norm{\rho - \sigma}_1$ is a more sensitive metric than fidelity --- trace distance measures the maximum distinguishability between states, whereas fidelity measures the distinguishability along the axis of the qubit state.   

To demonstrate $\SQM$ as a state stabilization technique, we prepare the qubit in the $\ket{0}$ state and induce coherent errors of increasing strength (Fig 2b). We plot $ 1-T(\rho, \sigma)$ with (grey squares) and without (green circles) $\SQM$ (Fig. 2b and 2d), overlayed with theoretical values from simulations performed using the measured initial state and no free parameters.  For small values of the error strength $\theta$, the corrected trace distance curve is nearly flat, indicating a first-order insensitivity to coherent errors. 

To use $\SQM$ to reduce the effects of deleterious coherent errors, one must know the axis along which to `measure’. In the single-qubit case, this corresponds to knowing the axis of the ideal qubit state --- in which case, the qubit can only store classical information. Stabilizer codes, on the other hand, are carefully designed so that parity measurements may be performed without disrupting the quantum information. When $\SQM$ is applied within the framework of a stabilizer code, coherent errors that would have moved information outside of the codespace are projected into incoherent errors. This both reduces the trace distance and prevents the build-up of coherent errors.   

To demonstrate the power of stabilizer measurement emulation, we use a simple code consisting of Bell states and stabilized by $ZZ$, with $\ket{\tilde{1}} = \frac{1}{\sqrt{2}} \ket{00} + \ket{11}$ and $\ket{\tilde{0}} = \frac{1}{\sqrt{2}} \ket{00} - \ket{11}$. To mitigate coherent errors for this code, we will apply $\SQM_{ZZ}$ as outlined n Eq. 5; namely, we apply identity gates or simultaneous $Z$ gates on both qubits with $50\%$ probability. Note that extending $\SQM$ to multiple qubits does not consume more time or require more complicated gates: with $p = 0.5$ we perform an identity operation, and with $p = 0.5$ we apply a $Z$ gate to each qubit. As with the state stabilization experiment, we apply a coherent error of varying strength (now an $R_x(\theta) \otimes R_x(\theta)$ error), either do nothing or apply $\SQM$, and measure the resulting trace distance to the initial state with a state tomography measurement (Fig 2b, 2c).  Here too, applying $\SQM$ leads to a distinctly smaller error. Though this plot demonstrates improved performance specifically for an $R_x(\theta) \otimes R_x(\theta)$ error, that particular error was chosen arbitrarily; any error that rotates the state outside of the ${\ket{\tilde{0}}, \ket{\tilde{1}}}$ manifold will behave similarly.

\section*{Using $\SQM$ to Mitigate Coherent Errors from 2QB Gates}

\begin{figure}[!ht]
	\center
	\includegraphics[width=\columnwidth]{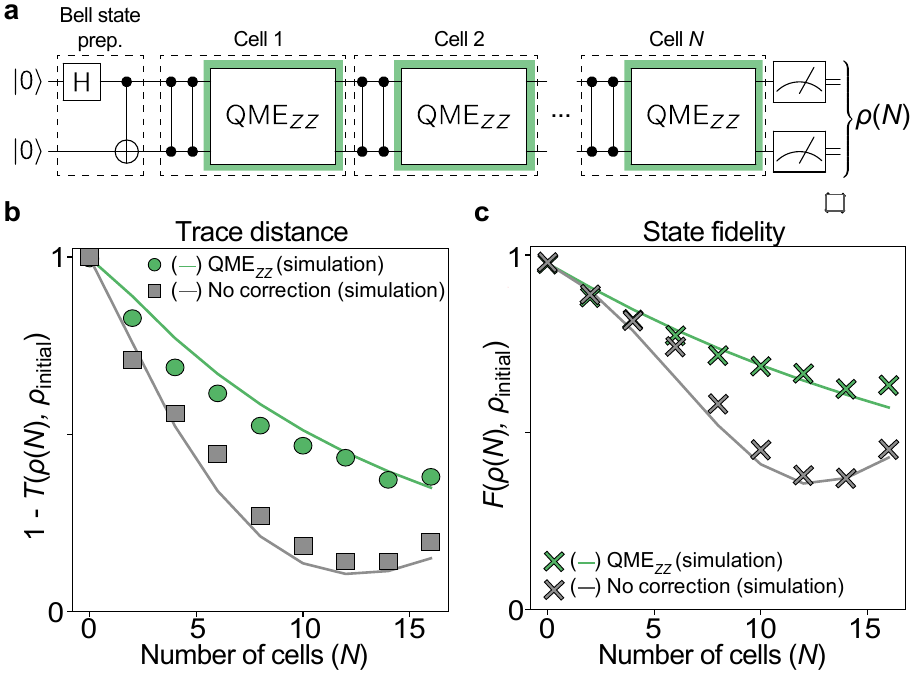}
		\caption{
			\footnotesize{
				\textbf{$\SQM$ for Mitigating Coherent Two-Qubit Errors in a Sequence of CZ Gates}.
				\textbf{a} Circuit used to test $\SQM$ in the context of a computation, consisting of a sequence of $\Id$ operations, each decomposed into two $\CZ$ gates. $\SQM_{XX}$ is added after every $\CZ$ pair. This circuit is run increasing the length of the sequence, and state tomography is used to reconstruct the trace distance and fidelity to the initially prepared state. The experiment is performed with a $\CZ$ gate with an intentionally introduced coherent error (\textbf{b}) and with a high fidelity $\CZ$ gate (\textbf{c}). To generate the theory curves we model the circuit with amplitude-damping and phase-damping channels, as well as single- and two-qubit over-rotations in the $\CZ$ gates. The amplitude (phase) damping constants are determined from measured $T_1$ ($T_2$) times, and the over-rotation angles are fit to the data. (In these plots, $1-T(\rho, \rho_{initial})$ has been normalized to 1 in order to account for SPAM errors.)
			} 
		}
	\label{fig:fig3}
\end{figure}

Two-qubit gates are the dominant source of coherent errors in all current generation quantum processors, so we explore the effects of $\SQM$ in a circuit comprised of a sequence of $\CZ$ gates (Fig. 3a). For this experiment we use a different code, stabilized by $XX$, with $\ket{\tilde{1}} = \frac{1}{\sqrt{2}} \ket{00} + \ket{11}$ and $\ket{\tilde{0}} = \frac{1}{\sqrt{2}} \ket{01} + \ket{10}$. ($\CZ$ is not a logical operation in this code, but $\CZ^2 = \Id \otimes \Id$ is a logical operation in every code.) To set a performance baseline, we prepare the  $\ket{\tilde{0}}$ state, apply $N$ $\CZ$ pairs and record the trace distance (state fidelity) to the initial state as shown in Fig3b(c). We then introduce error mitigation in the form of an $\SQM_{XX}$ inserted between every pair of $\CZ$ gates. 

Executing the circuit without $\SQM$ results in the steep plunge and revival that is the hallmark of the accumulation of coherent errors, seen in both the trace distance and the fidelity. The device used in this experiment has $\CZ$ gates with an average fidelity of 0.997 (measured via interleaved Clifford randomized benchmarking). Nevertheless, after 10 steps (20 gates) the output fidelity is nearly 0.5 --- a clear demonstration of the deleterious accumulation of coherent errors. Adding $\SQM$ leads to a smooth monotonic decay, which indicates that the coherent error has been made incoherent. Though adding $\SQM$ increases the circuit depth, it results in a substantial improvement in circuit performance.

\section*{Outlook}\label{sec:conclusion}
We have demonstrated a new feed-forward control technique for mitigating coherent errors in quantum information processing which is tailored to discrete coherent errors. By using stochastically-applied single-qubit gates to `emulate' quantum measurement along the appropriate axis, coherent errors can be made incoherent. Unlike randomized compiling, $\SQM$ does not require the computation of correction gates needed in randomized compiling, with the trade-off that it only protects against errors that rotate the qubit out of the logical codespace. $\SQM$ also offers protection against coherent errors that might have occurred during twirling gates. We show how $\SQM$ can be used in the context of stabilizer codes to improve circuit performance in terms of both trace distance and fidelity. This still holds in arbitrarily generated circuits where simple dynamical decoupling schemes do not offer an advantage. $\SQM$ is a promising addition to the collection of quantum computing control techniques.      

\begin{acknowledgments}
	\textbf{Acknowledgments:} 
 AG acknowledges funding from the 2019 Google US/Canada PhD Fellowship in Quantum Computing. 
 MK acknowledges support from the Carlsberg Foundation during part of this work.
IM acknowledges funding from NSF grant FET-1910859.
This research was funded in part by the U.S. Army Research Office Grant W911NF-18-1-0411 and the Assistant Secretary of Defense for Research \& Engineering under Air Force Contract No. FA8721-05-C-0002.
Opinions, interpretations, conclusions, and recommendations are those of the authors and are not necessarily endorsed by the United States Government.
\\

\textbf{Competing interests:} The authors declare no competing interests.
\end{acknowledgments}

\bibliography{biblio}{}
\bibliographystyle{science}

\end{document}


\title{Supplementary material for ``\thetitle''}
	
	\author{A.~Greene}\affiliation{\RLEaffil}
	\author{M.~Kjaergaard}\affiliation{\RLEaffil}
	\author{M.~E.~Schwartz}\affiliation{\LLaffil}
	\author{G.~O.~Samach}\affiliation{\RLEaffil}\affiliation{\LLaffil}
	\author{A.~Bengtsson}\affiliation{\RLEaffil}\affiliation{\CHALMERSaffil}
	\author{C.~McNally}\affiliation{\RLEaffil}
	\author{M.~O'Keeffe}\affiliation{\LLaffil}
	\author{D.~K.~Kim}\affiliation{\LLaffil}
	\author{M.~Marvian}\affiliation{\RLEaffil}\affiliation{\MITMECHaffil}
	\author{A.~Melville}\affiliation{\LLaffil}
	\author{B.~M.~Niedzielski}\affiliation{\LLaffil}
	\author{A.~Veps\"{a}l\"{a}inen}\affiliation{\RLEaffil}
	\author{R.~Winik}\affiliation{\RLEaffil}
	\author{J.~Yoder}\affiliation{\LLaffil}
	\author{D.~Rosenberg}\affiliation{\LLaffil}
	\author{S.~Lloyd}\affiliation{\RLEaffil}\affiliation{\MITMECHaffil}
	\author{T.~P.~Orlando}\affiliation{\RLEaffil}
	\author{I.~Marvian}\affiliation{\DUKEaffil}
	\author{S.~Gustavsson}\affiliation{\RLEaffil}
	\author{W.~D.~Oliver}\affiliation{\RLEaffil}\affiliation{\LLaffil}\affiliation{\MITPHYSaffil}\affiliation{\EECSaffil}

	\maketitle
	
	\section{Device} 
	
	\begin{figure}[!h]
	\label{supp:fig_device_schematic}
	\includegraphics[scale=.5]{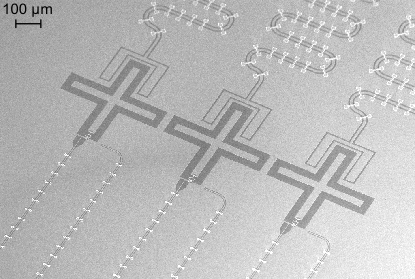}
	\caption{}
	\end{figure}

	The device used for these measurements is the same used for [34]. As shown in the SEM image in Supplementary Figure 1, this chip has three asymmetric `xmon'-style transmon qubits with individual microwave and flux control lines. All experiments in this paper were performed with the two leftmost qubits; their properties are reported in Table I. Interleaved randomized benchmarking data for single- (two-) qubit gates are shown in Supplementary Figure 2 (3). 
	
	\begin{table}[h]
		\label{supp:tab_device_properties}
		\centering
		\begin{tabular}{|c|c|c|}
			\hline
			& Qubit 1               & Qubit 2            \\
			Parameter       & ($\sigma$, target)    & ($\rho$, instruction)   \\
			\hline
			\hline
			Idling frequency, $\omega_i/2\pi$  & 4.748 GHz&4.225 GHz\\
			Anharmonicity, $\eta/2\pi$ & $-175$ MHz & $-190$ MHz\\
			Coupling strength, $g/2\pi$ & \multicolumn{2}{c|}{10.6 MHz\quad\quad\quad}  \\
			Readout resonator frequency, $f_i/2\pi$ &7.251 GHz& 7.285 GHz\\
			Junction asymmetry & 1:5 & 1:10\\
			\hline
			\hline
			Relaxation time at idling point, $T_1$ &  23 $\mu$s&  39 $\mu$s\\
			Coherence time at idling point, $T_{2\text{R}}$  &  13 $\mu$s&  25 $\mu$s\\
			
			Effective relaxation time undergoing \CZ{} trajectory, $\widetilde T_{1}$ $^1$ & $\approx 17 $ $\mu$s & (same as idling) \\
			Effective coherence time undergoing \CZ{} trajectory, $\widetilde T_{2\text{R}}$ $^1$ & $\approx 5$ $\mu$s & (same as idling)  \\
			\hline
			\hline
			Single-qubit gate fidelity, $t_\text{1qb}$ $^1$ & \textbf{\textit{\underline{}}} $\geq$ 0.999 & $\geq$ 0.999 \\
			Single-qubit gate time, $t_\text{1qb}$& 30 ns & 30 ns \\
			Two-qubit gate fidelity, $t_\text{\CZ}$ $^1$ & \multicolumn{2}{c|}{0.997\,\,\qquad\qquad} \\
			Two-qubit gate time, $t_\text{\CZ}$  & \multicolumn{2}{c|}{60 ns\,\,\qquad\qquad} \\
			\hline

		\end{tabular}
		\\ $^1$ See M. K. \textit{et al.} \textit{arXiv:2001.08838} (2020) for details
		\caption{Parameters of the two qubits used in this work.}

	\end{table}

	\section{State Stabilization Along An Arbitrary Axis} 
	\begin{figure}[!h]
		\label{supp:fig_arbitraryAxis}
		\includegraphics[width=1\columnwidth]{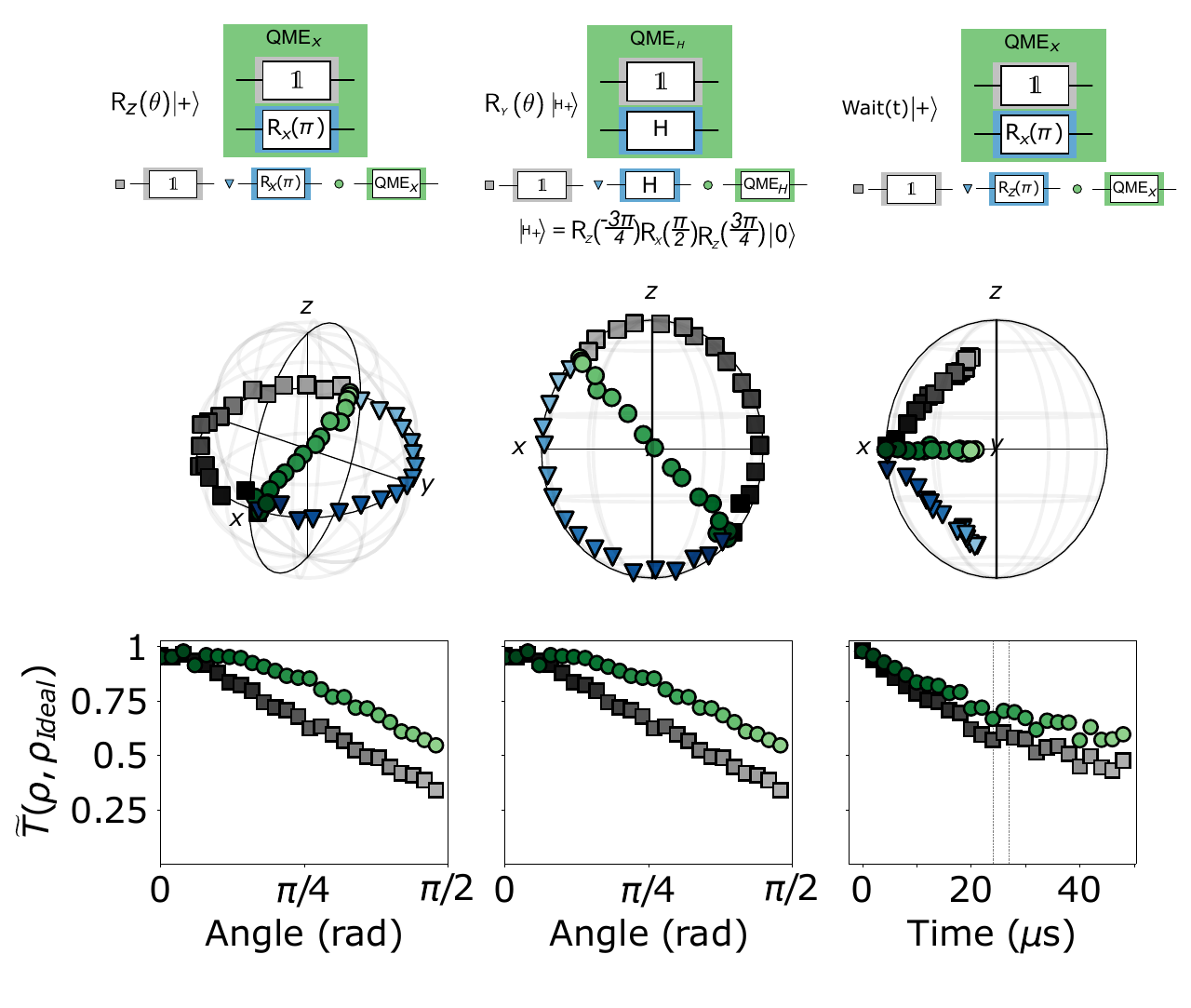}
		\caption{}
	\end{figure}

	In the main text, Equation 1 and Equation 2  describe how the quantum channel for measuring the expectation value of an arbitrary unitary operator $S_v$ can be performed by stochastically applying $S_v$. Figure 1 demonstrates this equivalence for a measurement along the $z$-axis. Here, we show similar data for emulated measurement along the $x$-axis (Supplementary Figure 4a) and along the $\frac{x+y}{2}$-axis (Suppementary Figure 4b).

	\section{Error Mitigation} 
	In Figure 2 of the main text, we demonstrate $\SQM$ as an error mitigation technique using an intentionally applied $\sigma_x \otimes \sigma_x$ over-rotation error of varying strength. In Figure \ref{supp:fig_arbitraryAxis} we show that $\SQM$ leads to first-order insensitivity to a variety of transversal errors. 
	
	\begin{figure}[!ht]
	\label{supp:fig_arbitraryAxis}
	\includegraphics[width=1\columnwidth]{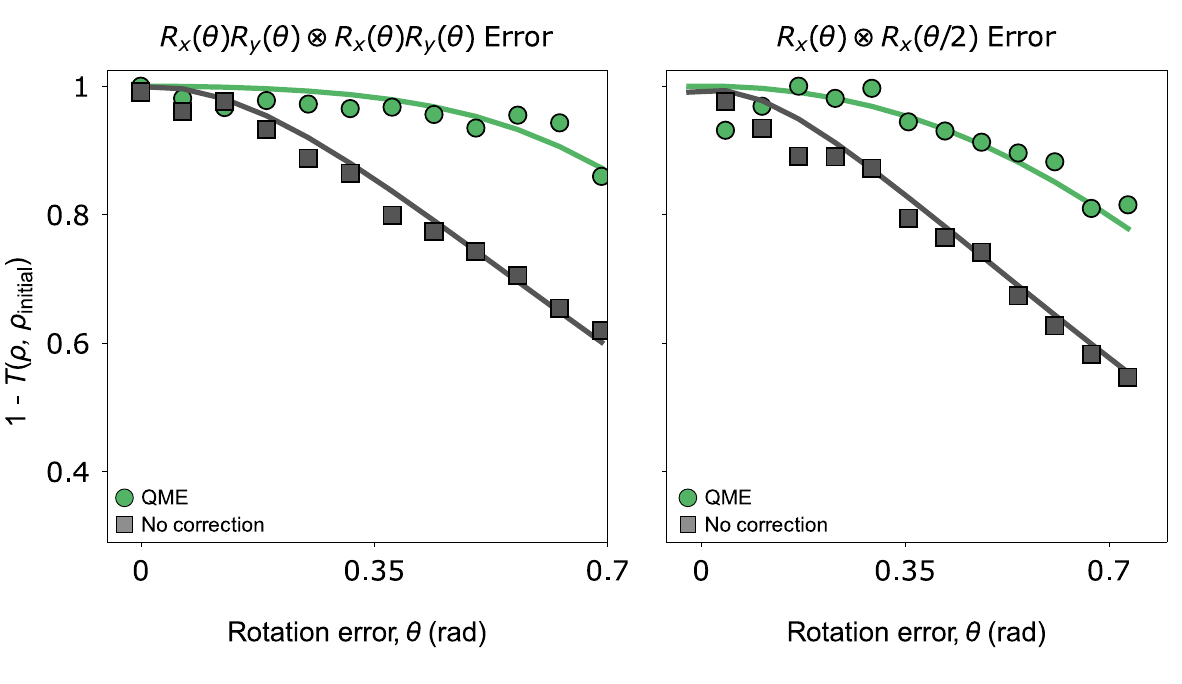}
	\caption{}
	\end{figure}
	
	\section{Simulations} 
	All theory curves used in this paper were generated by simulating the described circuit in \texttt{QuTiP}, using the initial $\rho$ as measured by state tomography. In Fig 3 and Fig 4, the nonidealities in the $\CZ$ gates were captured by representing each $\CZ$ gate as the composition of the ideal operation with a \textsf{CPHASE($\phi$)} over-rotation, single-qubit $Z$ over-rotations $R_Z(\theta_1) \otimes R_Z(\theta_2)$ and a leakage channel parameterized by leakage rate $lambda$. The parameters $\phi, \theta_1, \theta_2$ and $\lambda$ were fit to the state tomography data taken at each step of the circuit. The simulations for Fig. 3 and Fig. 4 also included decoherence, in the form of amplitude-damping and dephasing channels added after each gate in the circuit. The decay rates for these channels were determined by measured coherence times (shown in Table I) and the duration of the preceding gate (30n single-qubit gates and 60ns two-qubit gates, with a 5ns gap between pulses). 
	
	Our state tomography analysis does not discriminate between $\ket{1}$ and $\ket{2}$, so leakage from $\ket{11}$ to $\ket{20}$ looks like incoherent population transfer to $\ket{10}$. Accordingly, we used the following Kraus operators used to model leakage:
	
\begin{eqnarray}
&&  L_{1,\lambda}(t) = \begin{pmatrix}
1 & 0 & 0 & 0 \\
0 & 1 & 0 & 0 \\
0 & 0 & 1 & 0 \\
0 & 0 & 0 & \sqrt{1-\lambda} \\
\end{pmatrix},\quad
L_{2, lambda}(t) = \begin{pmatrix}
0 & 0 & 0 & 0 \\
0 & 0 & 0 & 0 \\
0 & 0 & \sqrt{\lambda} & 0 \\
0 & 0 & 0 & 0
\end{pmatrix} 
\label{eq-sup:amplitude_damping}
\end{eqnarray}

	The channel that composes amplitude damping and dephasing is given by:
	
\begin{equation}
\begin{gathered}
  \mathcal{E}_{\mathrm{q}k}(t): \rho_{\text{q}k} \mapsto \sum_{\substack{i = 1, 2 \\ j = 1, 2, 3}}
   A_{i,\Gamma_1}(t)  D_{j, \Gamma_\phi}(t) \rho_{\mathrm{q}k}  D^\dagger_{j, \Gamma_\phi}(t)  A^\dagger_{i, \Gamma_1}(t),\\
\end{gathered}
\end{equation}
where $A_{i, \Gamma_1}(t)$ is the amplitude damping process (with $\Gamma_1 = 1/T_1$), and $D_{j, \Gamma_\phi}(t)$ is the dephasing process ($\Gamma_\phi = 1/T_{2\text{R}} - 1/2T_1$), $\Gamma_{1, \text{q}k}$ and $\Gamma_{\phi, \text{q}k}$ are the appropriate coherence parameters for qubit $k$, and $t$ is the time of the preceeding single- or two-qubit gate on that qubit. The amplitude damping and dephasing Kraus operators are given by: 

\begin{eqnarray}
&&  A_{1,\Gamma_1}(t) = \begin{pmatrix}
  1 & 0 \\
  0 & e^{-\Gamma_{1,\mathrm{q}k} t / 2}
  \end{pmatrix}\quad
  \\
 && A_{2, \Gamma_1}(t) = \begin{pmatrix}
  0 & \sqrt{1 - e^{-\Gamma_{1,\mathrm{q}k} t}} \\
  0 & 0
  \end{pmatrix} \label{eq-sup:amplitude_damping}
  \\  
&& D_{1,\Gamma_\phi}(t) = \begin{pmatrix}
  e^{-\Gamma_{\phi,\mathrm{q}k} t / 2} & 0 \\
  0 & e^{-\Gamma_{\phi,\mathrm{q}k} t / 2}
\end{pmatrix}\quad
\\
&& D_{2,\Gamma_\phi}(t) = \begin{pmatrix}
  \sqrt{1 - e^{-\Gamma_{\phi,\mathrm{q}k} t}} & 0 \\
  0 & 0
\end{pmatrix}\quad
\\
&& D_{3,\Gamma_\phi}(t) = \begin{pmatrix}
  0 & 0 \\
  0 & \sqrt{1 - e^{-\Gamma_{\phi,\mathrm{q}k} t}}
\end{pmatrix}
\end{eqnarray}